\documentclass[prl,twocolumn,showpacs]{revtex4}
\usepackage{graphicx}

\begin{document}

\title{Symmetric Autocompensating Quantum Key Distribution}

\author{Zachary~D.~Walton}
\email{walton@bu.edu} \homepage[Quantum Imaging Laboratory
homepage:~]{http://www.bu.edu/qil}

\author{Alexander~V.~Sergienko}

\author{Lev B. Levitin}

\author{Bahaa~E.~A.~Saleh}

\author{Malvin~C.~Teich}
\affiliation{Quantum Imaging Laboratory, Department of Electrical
\& Computer Engineering, Boston University, 8 Saint Mary's Street,
Boston, Massachusetts 02215-2421}


\begin{abstract}
We present quantum key distribution schemes which are
autocompensating (require no alignment) and symmetric (Alice and
Bob receive photons from a central source) for both polarization
and time-bin qubits.  The primary benefit of the symmetric
configuration is that both Alice and Bob may have passive setups
(neither Alice nor Bob is required to make active changes for each
run of the protocol). We show that both the polarization and the
time-bin schemes may be implemented with existing technology. The
new schemes are related to previously described schemes by the
concept of advanced waves.
\end{abstract}

\maketitle

Of all the capabilities afforded by quantum information
science~\cite{Nielsen01}, quantum key distribution (QKD; for a
review, see Ref.~\cite{Gisin02}) currently shows the most promise
for practical implementation.  Accordingly, there has been a
concerted effort to develop QKD schemes that mitigate the
technical challenges associated with existing approaches.  Among
the successes in this effort are the development of
autocompensating (alignment-free) schemes for
polarization~\cite{Boileau03} and
time-bin~\cite{Muller97,Bethune98,Walton02a,Walton03a} qubits. A
further advance is the development of a symmetric scheme for
time-bin qubits in which neither Alice nor Bob is required to make
active changes to their setups~\cite{Brendel99}. Here we use the
term symmetric to describe QKD schemes in which a central source
distributes some number of photons to both Alice and Bob, such
that they share entanglement. This is in contrast to round-trip
and one-way configurations, in which the photons move according to
Bob$\rightarrow$Alice$\rightarrow$Bob and Alice$\rightarrow$Bob,
respectively.  In this Letter, we show that symmetry and
autocompensation can be combined in a single implementation, for
both polarization and time-bin qubits.

This Letter is organized as follows.  Beginning with
polarization-coded QKD, we first present a round-trip scheme in
which autocompensation is achieved by sampling the channel
birefringence twice (once on the way from Bob to Alice and once on
the way back). Second, we show how Klyshko's ``advanced wave
interpretation'' (AWI)~\cite{Belinsky92} can be used to transform
this round-trip scheme into a one-way scheme imbued with passive
detection. Third, we apply the AWI again to obtain a symmetric
autocompensating scheme in which both Alice and Bob have passive
setups.   We then repeat these three steps for time-bin-coded QKD.
Finally, we describe feasible implementations of the symmetric
autocompensating schemes for both polarization and time-bin
qubits.

\begin{figure*}
\includegraphics[scale=.9]{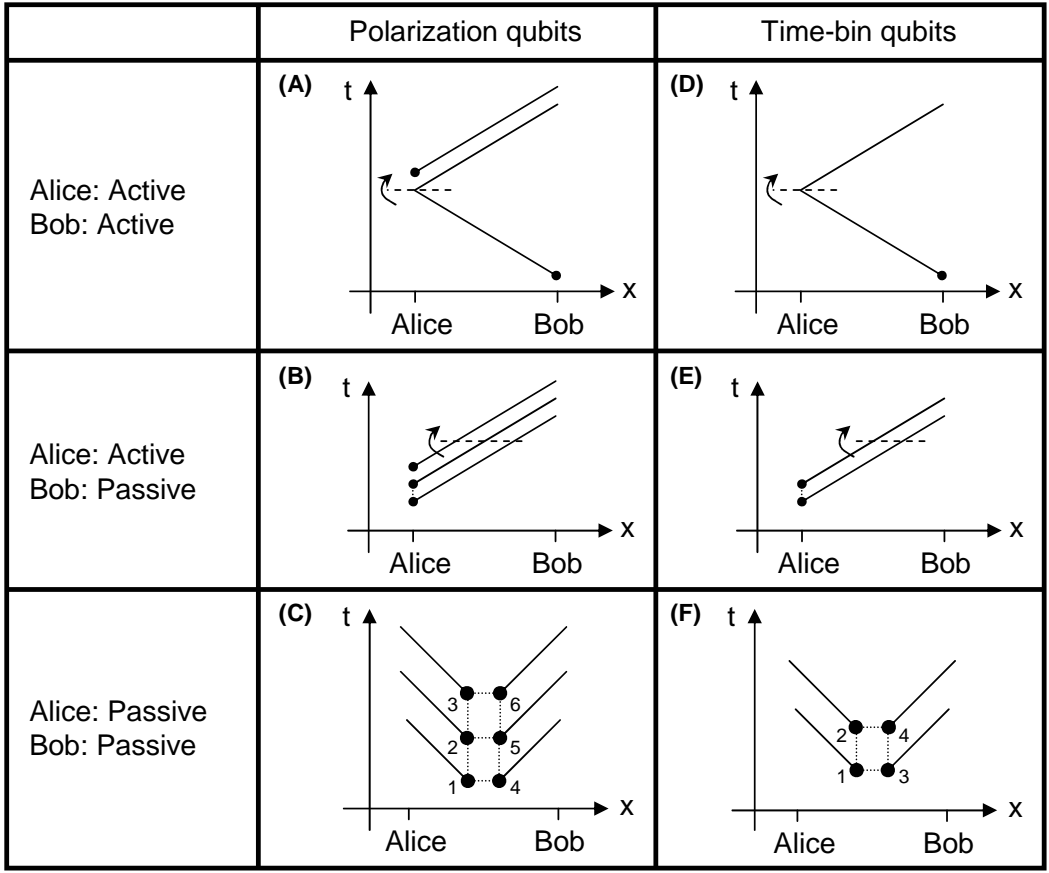}
\caption{Space-time diagrams of six autocompensating QKD schemes
organized by encoding (polarization or time-bin) and degree of
passivity.  The dashed lines and curved arrows show how the AWI
relates the round-trip schemes [(A) and (B)] to the one-way
schemes [(B) and (C)], and the one-way schemes to the symmetric
schemes [(C) and (F)]. The dotted lines connecting photons
indicate entanglement. The photon labels in (C) and (F) are used
later in this Letter.}\label{chart}
\end{figure*}

The left column of Fig.~\ref{chart} shows the space-time diagrams
of three autocompensating polarization-coded QKD schemes.  For
polarization qubits, autocompensating means that the scheme is
immune to channel birefringence.  The first scheme
(Fig.~\ref{chart}A) requires a round trip and is active (both
Alice and Bob are required to make changes to their respective
setups).  The scheme runs as follows.  Bob randomly chooses
between polarization states $|V\rangle$ and $|H\rangle+|V\rangle$
(here, and for the rest of this Letter, we suppress normalization
factors), and sends a single photon in that state to Alice. Alice
uses a Faraday mirror to reflect that single photon back, and also
sends along an auxiliary photon in the state $|V\rangle$. Alice
encodes a single bit by controlling the time ordering of the two
photons she sends to Bob. Bob then measures each photon in the
basis associated with the state of the initial photon he sent.
Without knowing which state Bob sent to Alice, Eve cannot
deterministically learn Alice's bit setting. From Bob's point of
view, the scheme is equivalent to Bennett's two-state
protocol~\cite{Bennett92c}, since he is attempting to
probabilistically distinguish between two nonorthogonal states.
The autocompensating feature is derived from the unique property
of the Faraday rotator: whatever the polarization transformation
along the line from Bob to Alice, the photon that Alice reflects
will arrive in Bob's lab in a polarization state orthogonal to its
original state~\cite{Martinelli89}.


The AWI was originally conceived as a method for generating
one-photon experiments from two-photon experiments.  However, we
may reverse this procedure and determine which two-photon state
embodies the action of Alice's Faraday rotator.  Using Faraday
rotation as an example, the AWI associates the single-photon
transformation
\begin{equation}\label{inout}
H_{\mbox{in}}\rightarrow V_{\mbox{out}}\quad\quad
V_{\mbox{in}}\rightarrow H_{\mbox{out}}
\end{equation}
with the two photon state
\begin{equation}\label{two-photon}
|H_{\mbox{in}}V_{\mbox{out}}\rangle+|V_{\mbox{in}}H_{\mbox{out}}\rangle\,
.
\end{equation}
In going from Eq.~(\ref{inout}) to Eq.~(\ref{two-photon}), the
propagation direction for $H_{\mbox{in}}$ and $V_{\mbox{in}}$ is
reversed.  To preserve the handedness of the coordinate system,
one of the transverse directions must be reversed as well. This
may be accomplished by replacing $V_{\mbox{in}}$ with
$-V_{\mbox{in}}$.  Thus, we see that the AWI associates Faraday
rotation with the polarization singlet state
$|HV\rangle-|VH\rangle$.
%
%

\begin{figure*}
\includegraphics[scale=.9]{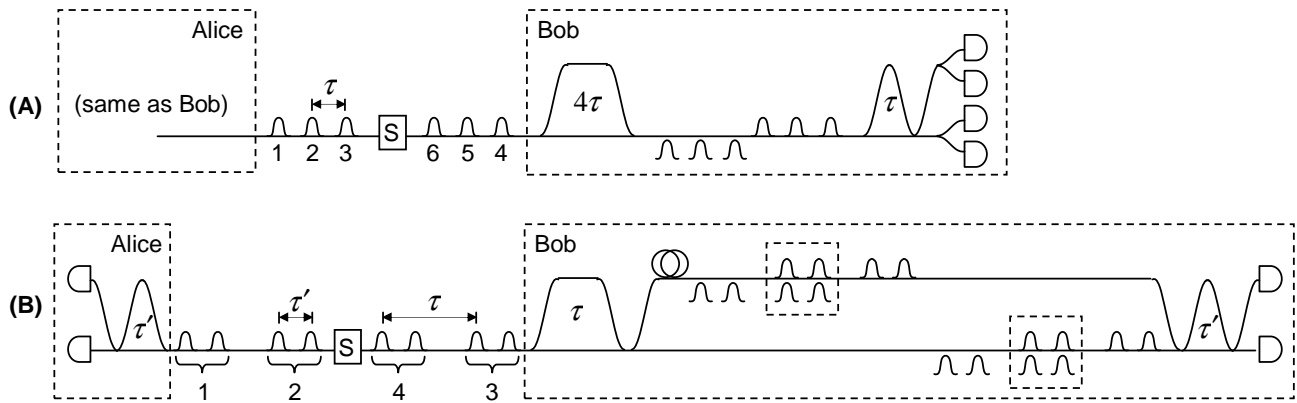}
\caption{Feasible implementations of the symmetric
autocompensating schemes of Figs.~\protect\ref{chart}C
and~\protect\ref{chart}F.  The schemes are symmetric in the sense
that both Alice and Bob receive photons from a common source (S);
they are passive in the sense that neither Alice nor Bob is
required to make active changes to their setups. For the
polarization case (A), the six photons are separately entangled
[see Eq.~(\protect\ref{pol})]. The three photons going to Bob are
sent through two Mach-Zehnder interferometers, the first with
relative delay $4\tau$, and the second with relative delay $\tau$.
The three photons below the line in Bob's apparatus are provided
as a visual indicator of the operation of the first
interferometer. For the time-bin case (B), the four photons are
separately entangled; however, all four photon are effectively
entangled by Bob's postselection of those occasions when one
photon is found in each of the small dotted boxes.}\label{impl}
\end{figure*}

We arrive at the one-way scheme of Fig.~\ref{chart}B by
``folding'' the input arm of the Faraday rotator of
Fig.~\ref{chart}A along the dashed line, thereby replacing a
round-trip single-photon space-time diagram with a one-way,
two-photon space-time digram (the dotted line connecting the two
photons indicates entanglement). What follows is a
passive-detection version of the three-photon scheme presented in
Ref.~\cite{Boileau03}. Alice sends three photons to Bob, with
either the first two (case 1), the last two (case 2), or the first
and last photons (case 3) in the singlet state, and the other
photon vertically polarized. Bob makes his measurements using the
passive setup shown on the right side of Fig.~\ref{impl}A.  By
appropriate postselection, this setup effectively makes a random
choice of two out of the three photons, and brings them together
on a non-polarizing beamsplitter, which serves to distinguish the
singlet state from the other three Bell
states~\cite{Braunstein95}. Ignoring the first Mach-Zehnder
interferometer (with relative delay $4\tau$) for the moment, we
see that the second interferometer (with relative delay $\tau$)
enables either the first two, or the last two, photons to meet at
the second beamsplitter of this interferometer.  If these two
photons are in the singlet state, they will leave by opposite
ports. The contrapositive is also true: if they leave by the same
port (and are detected by one of the pairs of detectors on each
output port), then one can infer that they were not in the singlet
state. Returning to the first interferometer, we see that this
interferometer provides an opportunity for the first and last
photons to be analyzed in a similar way. Thus, Bob's apparatus
probabilistically chooses a pair out of the three photons sent by
Alice, and determines whether the pair is in the singlet state or
in some orthogonal state.  Based on his detections, Bob can rule
out at most one of the three cases corresponding to Alice's
possible signal states. Therefore, after Bob has made his
detection, Alice announces whether the run was a ``data run''
(cases 1 or 2), or a ``test run'' (case 3).  The data runs are
used to share key material (one bit per run) and the test runs are
used to monitor the eavesdropper~\footnote{One can always convert
an active-detection scheme to a passive-detection scheme by using
a beamsplitter to probabilistically send the received photon(s) to
one of some number of separate detection setups.  A drawback of
this approach is that the number of optical elements required is
increased. The passive schemes described in this Letter, like that
in Ref.~\protect\cite{Brendel99}, are ``intrinsically passive,''
in that they achieve passive operation without increasing the
number of optical elements required.}.

We may apply the AWI one more time to get a six-photon symmetric
scheme (Fig.~\ref{chart}C) from the three-photon one-way scheme by
folding along the dotted line in Fig.~\ref{chart}B.  In this
scheme, the source produces the six-photon entangled state
\begin{eqnarray}\label{six}\nonumber
&&(\Psi^-V\Psi^-V-V\Psi^-V\Psi^-)_{123456}\\
&\equiv& |\Psi^-_{12}V^{\,}_3\Psi^-_{45}V_6\rangle -
|V_1\Psi^-_{23}V_4\Psi^-_{56}\rangle\, .
\end{eqnarray}
In Eq.~(\ref{six}) we use a compact notation that will simplify
expressions later in this Letter.  The execution of the protocol
is similar to the previous case, except that instead of randomly
choosing a three photon state and sending it to Bob, Alice uses
the same detection setup Bob uses (see Fig.~\ref{impl}A). By
inspecting the state in Eq.~(\ref{six}), we see that if Alice
determines that photons 1 and 2 are orthogonal to the singlet
state, then she knows that photons 5 and 6 are in the singlet
state. Similarly, if photons 2 and 3 are orthogonal to the
singlet, then photons 4 and 5 are in the singlet state.  Alice and
Bob can verify that the two terms in Eq.~(\ref{six}) are
coherently superposed (as opposed to statistically mixed) by
confirming that a certain joint detection (photons 1 and 3 in the
singlet state and photons 4 and 6 in the singlet state) never
occurs.  Since the singlet state is immune to collective
birefringence, this scheme, like the round-trip and one-way
schemes previously described, is autocompensating.

In the polarization case, only one of the schemes
(Fig.~\ref{chart}B) presented has been previously reported.  In
the time-bin case, the schemes in both Figs.~\ref{chart}D
and~\ref{chart}E have been described in Refs.~\cite{Muller97}
and~\cite{Walton02a,Walton03a}, respectively.  Therefore, we
immediately turn our attention to the symmetric time-bin scheme of
Fig.~\ref{chart}F.  The source produces the four-photon entangled
state
\begin{equation}\label{four}
(ELEL+LELE)_{1234}\, ,
\end{equation}
where $E$ and $L$ stand for early and late, respectively.  Alice
and Bob each have Mach-Zehnder interferometers with the delay
equal to the early/late time interval.  On the occasions when all
the early photons take the long path and all the late photons take
the short path, Alice and Bob announce their measurement results
and verify that the proper interference between the two terms in
Eq.~(\ref{four}) occurred.  On the occasions when at least one of
the photons on each side did not follow this
early$\rightarrow$long, late$\rightarrow$short pattern, Alice and
Bob are able to determine which of the terms in the superposition
was realized.  In this way they are able to share key material.
The scheme is passive because Alice and Bob simply record the time
of detection of single photons exiting the two output ports of
their respective Mach-Zehnder interferometers.  The scheme is
autocompensating because, on the occasions when interference
occurs between the two terms in Eq. (\ref{four}), each term picks
up the phase associated with two passes through each arm of both
interferometers.  Thus, the relative phase along the two paths of
each interferometer factors out and does not effect the measured
results.

It is clear from the states in Eqs.~(\ref{six}) and (\ref{four})
that the schemes of Figs.~\ref{chart}C and \ref{chart}F require
entangled states involving more than two particles.  Since the
direct generation of these states is not currently feasible, it is
important to determine if the schemes can be adapted to work with
some number of separately-entangled photon pairs.  This task is
particularly straightforward in the polarization case.  Using the
familiar notation for the four Bell states, and the mode labels in
Fig.~\ref{chart}C, we observe
\begin{eqnarray}\label{pol}\nonumber
&&\Phi^+\Phi^+\Phi^+_{142536}\\ \nonumber
&=&(\Psi^-\Psi^-+\Psi^+\Psi^++\Phi^-\Phi^-+\Phi^+\Phi^+)_{1245}\Phi^+_{36}\\
\nonumber
&=&(\Psi^-\Psi^-+\Psi^+\Psi^++\Phi^-\Phi^-+\Phi^+\Phi^+)_{2356}\Phi^+_{14}\\
&=&(\Psi^-\Psi^-+\Psi^+\Psi^++\Phi^-\Phi^-+\Phi^+\Phi^+)_{1346}\Phi^+_{25}\,
.
\end{eqnarray}
We can express this series of equations in words as follows. Take
three separately-entangled photon pairs (each pair in the state
$\Phi^+$), and, for each pair, send one photon to the left and the
other to the right.  Perform a Bell-basis measurement on any two
photons on the left, and the corresponding pair on the right will
collapse into whichever state results from the measurement of the
photons on the left.  Thus, Alice and Bob can replace the
six-photon entangled state of Eq.~(\ref{six}) with three pairs of
separately-entangled photon pairs, and implement a symmetric,
autocompensating protocol that is closely related to the one-way
scheme of Fig.~\ref{chart}B.  Specifically, whenever Alice detects
a pair of photons in the singlet state, she has effectively
prepared the corresponding pair of Bob's photons in the singlet
state.  This implementation can be seen as an application of
entanglement swapping~\cite{Zukowski93}.

Obtaining a feasible version of the time-bin implementation of
Fig.~\ref{chart} is also straightforward.  The setup in
Fig.~\ref{impl}B shows how Alice and Bob may implement this scheme
using separately-entangled photon pairs. Instead of the state in
Eq.~(\ref{four}), the source creates the state
$\Phi^+_{13}\Phi^+_{24}$;  and, instead of a simple Mach-Zehnder
interferometer, Bob has the detection setup shown in
Fig.~\ref{impl}B.  The first interferometer in Bob's setup allows
the first photon to meet the second photon at a non-polarizing
beam splitter.  By postselecting the occasions when one photon is
found to be in each of the small dashed boxes, Bob effectively
entangles the two photons sent to him in precisely the way
required by Eq.~(\ref{four}).  From this point, Alice and Bob each
analyze their photons with Mach-Zehnder interferometers, and the
scheme proceeds as previously described. This technique can be
viewed as the time-bin analog of the polarization-based
entanglement distillation experiment described in
Ref.~\cite{Yamamoto03}.

It is interesting to observe that discoveries in the field of
quantum information (entanglement swapping and entanglement
distillation) can be naturally related to other areas of quantum
information theory (quantum error correction and decoherence-free
subpaces) via the AWI, as demonstrated in Fig.~\ref{chart}.  Since
the central goal of quantum computation is a ``folding in time''
of a classical computation, the AWI may yield insight into the
mechanisms behind the speed-up achieved by certain quantum
algorithms.

We have presented symmetric autocompensating QKD schemes that can
be implemented with existing technology for both polarization and
time-bin qubits.  The primary benefit offered by these new schemes
is passive operation (neither Alice nor Bob is required to make
active changes to their setups).  While the schemes make use of
existing two-photon sources, it is important to point out that
current techniques for producing and detecting multiple photon
pairs have very low yields ($\sim$1 detected four-fold coincidence
per second~\cite{Zhao03}).

This work was supported by the National Science Foundation; the
Center for Subsurface Sensing and Imaging Systems (CenSSIS), an
NSF Engineering Research Center; and the Defense Advanced Research
Projects Agency (DARPA).


\end{document}